\documentclass{article}
\usepackage{spconf,amsmath,graphicx}
\usepackage{booktabs}
\pdfoutput=1
\title{Multitask-based joint  learning approach to robust ASR for radio communication speech}
\name{Duo Ma$^{12}$, Nana Hou$^2$, Van Tung Pham$^2$, Haihua Xu$^2$, Eng Siong Chng$^{23}$ }
\address{
$^1$ Human Language Technology (HLT) Laboratory,\\
Department of Electrical and Computer Engineering,\\
National University of Singapore, Singapore\\
$^2$School of Computer Science and Engineering, Nanyang Technological University, Singapore\\
$^3$Temasek Laboratories, Nanyang Technological University, Singapore\\
maduo25@163.com,{\{hhx502,tungphamvan\}}@gmail.com,\\NANA001@e.ntu.edu.sg,ASESChng@ntu.edu.sg}

\begin{document}
%{}
\maketitle

\begin{abstract}
To realize robust End-to-end Automatic Speech Recognition (E2E ASR) under radio communication condition, we propose a multitask-based method to jointly train a Speech Enhancement (SE) module as the front-end and an E2E ASR model as the back-end in this paper. One of the advantages of the proposed method is that the entire system can be trained from scratch, i.e.,  different from prior works, either component here doesn't need to perform pretraining and fine-tuning processes separately. Through analysis, we found that the success of the proposed method lies in the following aspects. First, multitask learning is essential, that is, the SE network is not only learned to produce more intelligible speech, it is also aimed to generate speech that is beneficial to recognition. Secondly, we also found speech phase preserved from noisy speech is critical for an improved ASR performance. Thirdly, we propose a dual-channel data augmentation training method to obtain further improvement. Specifically, we combine the clean and enhanced speech to train the whole system. %We evaluate the proposed method on the RATS English data set, achieving 2.5\% absolute Word Error Rate (WER) with the joint training method, and up to 6.1\% absolute WER with the proposed data augmentation method.
 We evaluate the proposed method on the RATS English data set, achieving a relative WER reduction of ~4.6\% with the joint training method, and up to a relative WER reduction of 11.2\% with the proposed data augmentation method.
\end{abstract}
\section{Introduction}
\label{sec:intro}
With the surge of recent attention-based end-to-end neural network modeling framework~\cite{chan2015listen,sainath2019two,he2019streaming,gulati2020conformer,li2021better}, as well as big data usage, the performance of Automatic Speech Recognition (ASR) has been significantly improved, such that its application has been widely deployed in diversified industrial area. However, ASR performance is still far from being desired under extremely noisy conditions, such as radio communication conditions, where speech might not only be contaminated by ambient noise, it is also distorted by communication channel due to limited transfer bandwidth, as well as Codec losses.
For instance, for radio communication speech \footnote{By radio communication speech, here it means single channel Ultra High Frequency (UHF) speech that is very noisy. The SNR is close to 0 dB.} to be studied in this work, it not only has lower Signal-to-Noise Ratio (SNR), the speech signal itself is also seriously distorted. As a result,  the speech intelligibility is rather low. 

To achieve decent results under noisy conditions, one common approach is to employ multi-condition
training method. This is appropriate for some  minor or intermediate noisy conditions. For the extremely noisy conditions, such as SNR being close to 0 dB, the first priority is to make the incoming speech intelligible. As a result,
Speech Enhancement (SE) as the front-end is necessary. Nevertheless, prior experiences tell us employing SE to boost speech intelligibility does not mean the enhanced speech is necessarily conducive to the back-end ASR performance improvement \cite{donahue2018exploring,pandey2021dual}, given that the SE and ASR models are trained separately. Besides, even both SE and ASR models are jointly trained, it is not guaranteed with improved ASR performance. This is particularly true under the single channel scenarios.

In this paper, we propose a multitask-based joint learning approach to robust ASR over radio communication speech, i.e., RATS \cite{graff2014rats} English data set. The entire network is a pipeline that is made up of an E2E SE and ASR components respectively, and the front-end SE component provides denoised speech for better ASR results in the back-end.
The so-called multitask-based joint learning approach refers to the front-end SE component is not only learned to produce more intelligible speech (whose loss is denoted as $\mathcal{L}_{SE}$), it is also learned to yield speech that boost ASR results (whose loss is denoted as $\mathcal{L}_{ASR}$). 

\begin{figure*}[!htb]
\centering
 \includegraphics[width=0.9\textwidth]{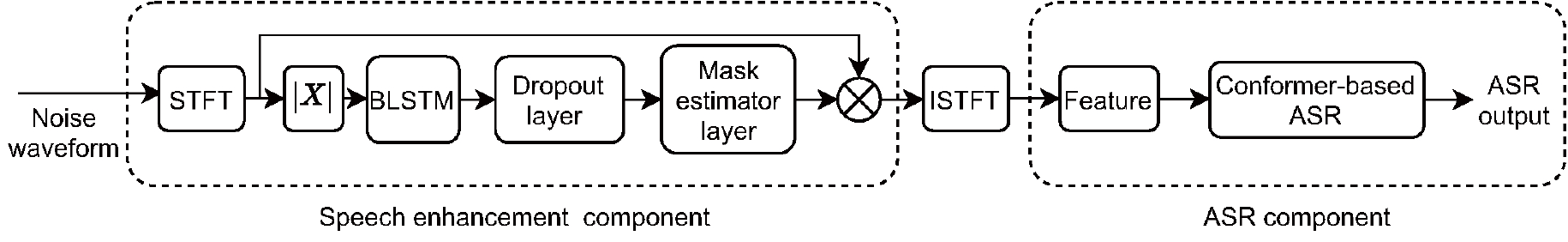}
\caption{Speech enhancement and ASR joint modeling architecture. We use  time-frequency spectral masking based method for SE, taking STFT spectral magnitude as input during training.}
\label{fig:joint-modeling-arch}
\end{figure*}

The main contributions of this paper are four-fold. First, the entire ASR system can be simply trained from scratch with multitask-based joint learning approach. 
Secondly, we have performed comprehensive analyses on how ASR performance is affected by the front-end SE component.
Particularly, the total loss function for the SE component is defined as %$\mathcal{L}_{ASR} + \alpha\mathcal{L}_{SE}$, 
{$(1-\beta)\mathcal{L}_{ASR} + \beta\mathcal{L}_{SE}$},
where $\beta$ is a scaling factor controlling the contribution of the SE component. Besides, we have explicitly verified that phase information is critical to ASR performance improvement. 
Thirdly, we propose a dual-channel data augmentation method using clean and enhanced speech to train the back-end ASR component, and it leads to further improvement.
Finally, to the best of our knowledge, our work is the first time report on robust ASR assisted with SE over radio communication speech.

%The paper is organized as follows. Section 2 introduces the related work. Section 3 and 4 describe the joint modeling architecture and the proposed multitask-based joint learning.
%In section 5, experimental settings and results are presented. Section 6 concludes the study.
The paper is organized as follows. Section ~\ref{sec:prio-work} introduces the related work. Section ~\ref{sec:joint-arch}
and ~\ref{sec:multitask} describe the joint modeling architecture and the proposed multitask-based joint learning.
In section ~\ref{sec:experiments}, experimental settings and results are presented.

\section{Related Work}~\label{sec:prio-work}
In earlier times, SE system is separately trained as the pre-processor for robust ASR systems.
However, the main challenge of separate training is that the output from the SE system is actually a distorted speech that may not be desirable for the ASR. To resolve such a mismatch problem, prior work \cite{deblin-osu-icassp2018-03,peter-osu-arxiv2018-01} propose a mimic loss from ASR output in addition to the conventional feature-based MSE loss to train the SE system. 
\cite{keishuke-ntt-icassp2020-11} employs a convolutional time-domain audio separation network (Conv-TasNet), which is utilized in~\cite{yi-conv-tasnet-arxiv201-04} for single-channel speaker-independent speech separation. The success of Conv-TasNet might be attributed to that spectral and phase features are not decoupled for consideration, overcoming the limitation of some spectral mapping~\cite{wang2017maximum} or time-frequency masking methods~\cite{yu2017permutation}. Likewise, to preserve the phase information when doing SE, \cite{zhongqiu-taslp2020-16} proposes a complex spectral mapping for both single and multiple channel SE for robust ASR. The advantage is particularly demonstrated for the multi-channel ASR performance improvement.
Besides, to alleviate distortions generated by the target SE system, \cite{peidong-taslp2019-07} 
employs diversified noise data to train a distortion-independent SE system for robust ASR, which obtains decent WER results on ChiME-2 corpus.
Likewise, \cite{chris-google-icassp-2018-09} introduces SE Generative Adversarial Network (SEGAN). However, it is found that SEGAN brings  performance improvement only when the enhanced data is combined with the original data for training. 

Recently, joint training of SE and ASR systems become popular.
\cite{zhong-taslp2016-05} proposes a joint training framework for ChiME-2 task, where SE front-end is jointly trained with the conventional DNN-HMM hybrid ASR system.
In \cite{bin-tecent-ins2019-13} and \cite{lujun-tumg-arxiv2021-06},  SEGAN assisted joint training methods are introduced for the AISHELL-1 simulated-noise data respectively. Notably, to make SEGAN work, pretraining is indispensable for the generator part, and the fine-tuning of the generator with the corresponding discriminator is a non-trivial work. Different from \cite{bin-tecent-ins2019-13} and \cite{lujun-tumg-arxiv2021-06}, no pretraining for either component is necessary in our case.

\begin{figure*}[htb]
\centering
 \centering{\includegraphics[width=0.9\textwidth]{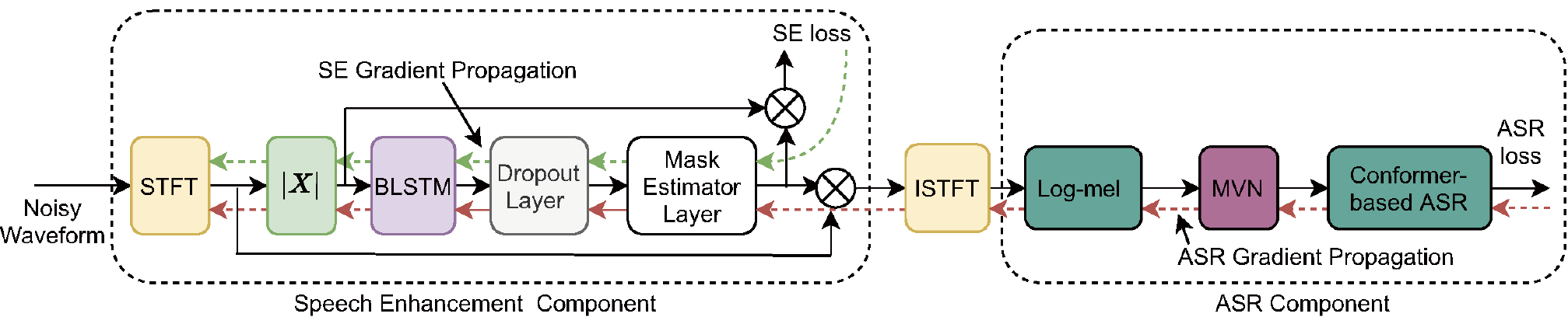}}
  \caption{Multitask-based joint learning framework for robust ASR over radio communication speech. Here, multitask learning means the front-end SE is updated by both SE and ASR back gradient propagation simultaneously.}
 \label{fig:multitask-joint-arch-01}
 %\vspace{-0.2in}
\end{figure*}

More recently, 
% assuming SE as front-end providing distorted speech for the back-end ASR system, 
\cite{ashutosh-fb-icassp2020-08} attempts to employ a SE-based DCCRN~\cite{hu2020dccrn} as data augmentation technique, using consistency loss to fine-tune the DCCRN component. To gain ASR improvement, a 3-step training recipe is employed. During decoding a learnable feature selection is adopted assuming that enhanced speech is complementary to the original speech.  Another work in~\cite{chanwoo-samsung-icassp2021-12} proposes a multi-channel-like data augmentation method, using SE as front-end. It aims to stablize an E2E-based streaming ASR in the back-end.
The model architecture is similar to one of our proposed methods, however, their training recipe is rather complicated.

\section{Joint Modeling Architecture}~\label{sec:joint-arch}
To realize robust ASR under noise condition, we propose the joint modeling architecture, which consists of two components, namely, the SE as front-end and the ASR in the back-end.  The SE front-end aims to provide enhanced speech conducive to the back-end ASR. Since we attempt to train the two components jointly, we concatenate them in tandem, as illustrated in
Figure~\ref{fig:joint-modeling-arch}. The fundamentals for each component are briefly described in this section.

\subsection{Time-frequency masking-based SE}~\label{sub:tfm-se}
As shown in Figure~\ref{fig:joint-modeling-arch}, we employ Bidirectional Long Short Term Memory (BLSTM) to conduct time-frequency masking-based SE work similar to~\cite{yu2017permutation}. Specifically, we use spectral magnitude $\lvert X \rvert$ extracted by short-time Fourier transform~(STFT) from the noisy waveform as input to train LSTM-based mask estimator $M$, where $X$ is complex Fourier transform, and $X=\mathcal{R} + i\mathcal{I}$. 
Such predicted masks $M$ conduct the element-wise product with the noisy input $X$, and then inverse STFT (ISTFT) transforms enhanced waveform from the corresponding features, that is, 
$\hat{X} = \textrm{ISTFT} ((\mathcal{R} + i\mathcal{I}) \otimes M)$.  
 
\subsection{Conformer-based ASR}~\label{sub:conformer-asr}
We use a Conformer-based ASR~\cite{gulati2020conformer,guo2021recent} for the back-end. However, we don't utilize SpecAugment~\cite{park2019specaugment}, because we find that it performs worse in our current experiment corpus. The Conformer is a convolution-augmented Transformer~\cite{vaswani2017attention}, based on the complementary features of convolutional learning and multi-head self-attention (MHSA). Therefore, the Conformer focuses more on feature locality, and is capable of learning global context dependencies. In practice, a proposed Conformer block takes the place of conventional Transformer block. \cite{gulati2020conformer} reported that the Conformer achieves consistent performance improvement over Transformer on Librispeech data set.
 
Except for the Conformer framework, our E2E ASR model is jointly trained with both CTC and attention-based cross-entropy criteria. As a result, the ASR loss criterion is as follows:

\begin{equation}~\label{eqn:asr-loss}
\begin{split}
     \mathcal{L}_{ASR}(Y|X_{enc}) &= (1-  \lambda)\mathcal{L}_{att}(Y|X_{enc}) \\
     &+ \lambda\mathcal{L}_{CTC}(Y|X_{enc}) 
\end{split}
\end{equation}
where $X_{enc}$ and $Y$ represent the encoder output and decoder output respectively. $\lambda \in [0, 1]$ aims to balance the losses between the CTC and attention-based cross-entropy criteria. For simplicity, we fix $\lambda=0.3$ during training.

\section{Multitask-based Joint Learning}
\label{sec:multitask}

\subsection{Single channel joint SE and ASR}
\label{sub:single-channel}

As shown in Figure \ref{fig:joint-modeling-arch}, we can use ASR loss in Equation \ref{eqn:asr-loss} to train the entire network jointly.
However, we found that such a training recipe yields
suboptimal results. This might be that our training data is too noisy, so ASR loss can not explicitly guide the SE component to denoise data. As shown in Figure \ref{fig:multitask-joint-arch-01}, we employ multitask-based joint training instead, where the front-end SE component is learned from both ASR and SE loss simultaneously. Here, the loss is defined as:
\begin{equation}
\label{eq:joint}
    \mathcal{L}_{joint} = (1-\beta)\mathcal{L}_{ASR} + \beta\mathcal{L}_{SE}
\end{equation}
where $\mathcal{L}_{ASR}$ is the ASR loss criterion defined in Equation~\ref{eqn:asr-loss}. $\mathcal{L}_{SE}$ is SE loss, where spectral magnitude MSE is employed. $\beta$ is the weighting factor that controls SE loss $\mathcal{L}_{SE}$.

As mentioned above in Section \ref{sec:prio-work}, multitask-based joint training as shown in Figure \ref{fig:multitask-joint-arch-01} has also been proposed previously. However, to the best of our knowledge, no work has emphasized how the output of SE is concatenated with ASR as input in detail. Actually, there are two options worth for our attention. 

\subsection{Preserve phase information}
\label{sub:phase}

Assuming $X^{\prime}$ is an enhanced complex spectrum, we can choose what follows as ASR input: 
\begin{enumerate}
\item $\lvert X^{\prime} \rvert \to \textrm{Log-Mel}({\lvert X^{\prime} \rvert}^2)$
\item $X^{\prime} \to \textrm{ISTFT}(X^{\prime}) \to X^{\prime\prime} \to \lvert X^{\prime\prime} \rvert \to \textrm{Log-Mel}({\lvert X^{\prime\prime} \rvert}^2)$ 
\end{enumerate}

In case 1, we ignore the phase information of the enhanced speech, while in case 2, we preserve the phase information of the enhanced speech. Moreover, case 2 is also more flexible, as $X^{\prime}$ and $X^{\prime\prime}$ can be in different dimensions. In this paper, we choose the second case (see Figure~\ref{fig:multitask-joint-arch-01}) but report the ASR results in both cases to indicate that phase information is decisive for WER improvements on our data. We note that joint training is viable for both cases.

\begin{figure*}[!htb]
\centering
 %\centerline{\includegraphics[width=0.5\textwidth]{Latex/figure/multitask-joint-learning-02.pdf}}
 \includegraphics[width=0.9\textwidth]{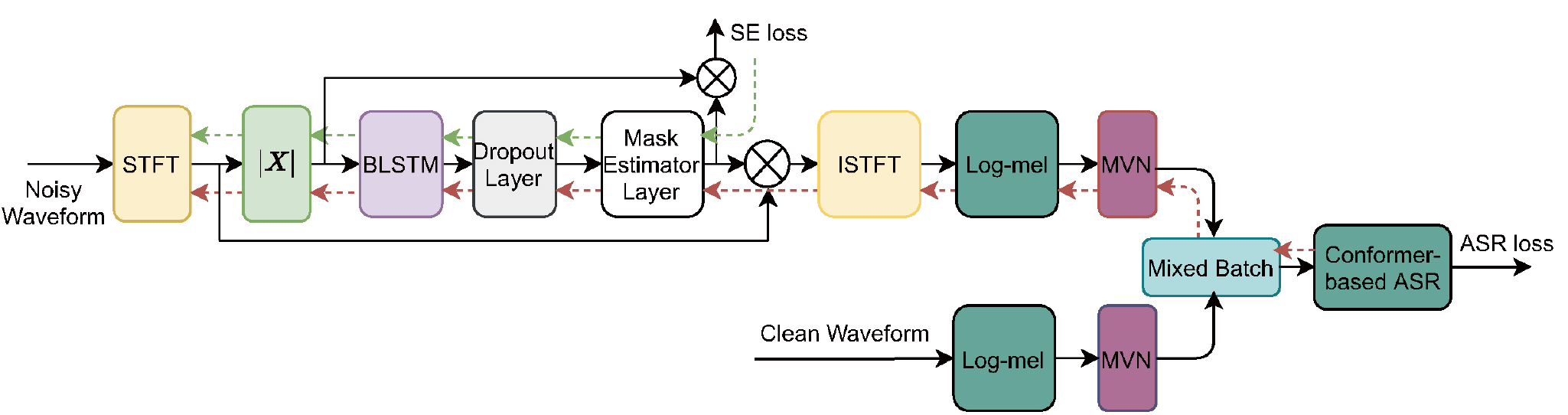}
\caption{Data flow diagram of dual-channel data augmentation for the  joint modeling architecture.}
\label{fig:multitask-join-arch-02}
\end{figure*}

\subsection{Dual-channel data augmentation}
\label{sub:two-channel}

Building a robust joint SE and ASR system under the radio communication speech conditions is a challenging problem, as the radio communication speech data is extremely noisy. The problem usually appears at initial training stage as the SE component cannot provide quality enhanced speech to the ASR component. As a result, the model may not be well learned in the end. To obtain robust ASR, we propose a dual-channel data augmentation method, as illustrated in Figure \ref{fig:multitask-join-arch-02}. 
Specifically, during the training, we mix the clean and enhanced speech in each mini-batch to train the joint network. We use the entire back-propagation to update the ASR component, while part of the back-propagation corresponding to the enhanced speech to update the SE component. During the testing, since we only have noise data, we use the same network as illustrated in Figure \ref{fig:multitask-joint-arch-01}. The ASR and SE losses are changed as follows:

\begin{eqnarray}
\label{eqn:joint}
\label{eqn:dual-loss}
\left \{
\begin{aligned}
    \mathcal{L}_{ASR}^{joint} =  \gamma\mathcal{L}_{ASR}^{C} +(1- \gamma) \mathcal{L}_{ASR}^{N} \\
    \mathcal{L}_{SE}^{joint}=\beta\mathcal{L}_{SE} +  (1-\beta)\mathcal{L}_{ASR}^{N} 
    \end{aligned}
    \right.
\end{eqnarray}
where $\mathcal{L}_{ASR}^{C}$ and $\mathcal{L}_{ASR}^{N}$ are defined as ASR loss from clean and noise data respectively. $\gamma$ is weighting factor for the clean data, while $\beta$ is the same as 
in Equation \ref{eq:joint}.

\section{Experiments and Results}
\label{sec:exp}
\label{sec:experiments}
\subsection{Data}
\label{sub:exp-data}
We conduct experiments on part of the English data that is originally utilized for Speech Activity Detection (SAD) from Robust Automatic Transcription of Speech (RATS) program over radio channels \cite{graff2014rats}. There are eight channels, and we choose channel A data that belongs to UHF data category for evaluation. The details are shown on the Table \ref{tab:overall-data}. The data is recorded with push-to-talk transceiver by playing back the clean Fisher data. One can refer to \cite{graff2014rats} for more details.

\begin{table}
\caption{The overall data sets (hours) for evaluation}
\label{tab:overall-data}      
\centering
\begin{tabular}{llll}
\toprule
% \hline\hline
% \hline\noalign{\smallskip}
Language & Train & Valid & Test  \\
% \hline\hline
% \noalign{\smallskip}\hline\noalign{\smallskip}
\midrule
 English & 44.3 & 4.9 & 8.2 \\
% \hline
% \noalign{\smallskip}\hline
\bottomrule
\end{tabular}
\end{table}

\subsection{Experimental setup}
\label{sub:exp-setup}
All experiments are performed on ESPnet~\cite{watanabe2018espnet} platform. We employ Adam algorithm \cite{kingma2014adam} to optimize the joint network as shown in the above figures with 0.002 and 32 as initial learning rate and batch size respectively.

\subsubsection{Speech Enhancement}
\label{subsub:se}
For SE component implementation as shown in Figure~\ref{fig:multitask-joint-arch-01} and \ref{fig:multitask-join-arch-02}, the network consists of 3 BLSTM layers with 896 units for each, then a dropout layer, and a feed-forward layer. The input to the BLSMT is 257-dimensional spectral magnitude features. To examine the effect of different masking estimate method, we also employ different activation functions, such as ReLU~\cite{nair2010rectified}, Mish~\cite{misra2019mish}, as well as meta-ACON~\cite{ma2020activate}which is  modified to fit sequence modeling ,Code is available at \footnote{https://github.com/shanguanma/joint-se-asr/blob/main/meta-acon.py} respectively.   

% In the SE implementation detailed in Figure \ref{fig:multitask-join-arch-02}, we utilized the noisy 257-dimensions magnitude features as the inputs. Then, the SE module predicted the masks for such noisy inputs by three BLSTM layers with 896 units, followed by a dropout (=0.5) and one feed-forward layer with 257 units. We explored three different activation functions for the last feed-forward layer including ReLU\cite{nair2010rectified}, Mish\cite{misra2019mish} and Meta-Acon \cite{ma2020activate}.

\subsubsection{Conformer-based ASR}
\label{subsub:asr}
We use Conformer \cite{guo2021recent} for the back-end ASR with 80-dim Log-Mel features as input. The encoder consists of 12 Conformer layers, while the decoder has 6 transformer layers, with 994 byte-pair-encoding (BPE) \cite{kudo-richardson-2018-sentencepiece} tokens as output. To yield better results, RNNLM-based shallow fusion \cite{karita2019comparative} is employed via training a RNNLM with the training transcript.

% In the part of ASR Module, output enhancement waveform is passed to front end of ASR module, the front end contains log-mel layer ,in our experiments, its dimension was 80. And global or utterance mean variance  normalization layer and forms 80 dimension representation, which is further subsampled by a factor of 4. The output respresentation is passed 12 Conformer\cite{guo2021recent} layer as encoder of asr Module, 6 transformer\cite{karita2019comparative} layer as decoder of ASR Module.All the joint SE and ASR model are trained to predict 994 byte-pair-encoding (BPE) tokens\cite{kudo-richardson-2018-sentencepiece}. The decoding is rescored with  RNNLM of BPE unit via shallow fusion \cite{karita2019comparative}.

\subsection{Results}
\label{sub:res}
Table \ref{tab:res} reports the overall WER results with both single- and dual-channel multitask-based joint learning (denoted as MTJL and DC-MTJL in Table \ref{tab:res}) methods respectively.

\begin{table}[ht]
\centering
\caption{WER (\%) comparison between different systems. MCT refers to multi-conditional training using 3x speed perturbation; Disjoint training refers to both SE and ASR systems are trained separately; JL stands for joint learning without multitask recipe, i.e., the entire network is learned from ASR loss; MTJL and DC-MTJL refer to single- and dual-channel multitask-based joint learning methods respectively.}
\label{tab:res}
\begin{tabular}{ccc}
\toprule
% \hline\hline
System &  Description & WER (\%) \\
% \hline\hline
\midrule
$\textrm{S}_1$ & Baseline with Global MVN  & 54.3  \\
$\textrm{S}_2$ & $\textrm{S}_1$, Speed perturbation (3x) & 49.8\\ 
% \hline
\midrule
$\textrm{S}_3$ & Disjoint training  & 55.6\\
%JL     & w/o phase and w/o multitask     &   xx \\
$\textrm{S}_4$ & JL (mono-task), with phase & 54.0 \\
$\textrm{S}_5$ & MTJL, $\beta=0.3$, w/o phase     &   68.6 \\
$\textrm{S}_6$ & MTJL, $\beta=0.3$ in Eq. (\ref{eq:joint}) & 51.8\\
$\textrm{S}_7$ & DC-MTJL, with  $\beta=0.3$ and $\gamma=0.7$ & \textbf{48.2}\\
% \hline
\bottomrule
\end{tabular}
\end{table}

It is note-worthy that both MTJL and DC-MTJL have achieved significant WER reduction compared with the baseline system in Table \ref{tab:res}. Particularly, the performance improvement of the proposed dual channel MTJL method is a relative reduction of 11.2\% over the baseline system. Additionally, one can notice that phase information is critical. Without phase consideration, the WER of System $\textrm{S}_5$ is rapidly degraded, while $\textrm{S}_4$ improved the performance with phase information for even mono-task based joint training. Furthermore, with the help of both phase information and multitask learning, System $\textrm{S}_6$ achieves significant WER improvement over the Baseline $\textrm{S}_1$, from 54.3\% down to 51.8\%. Thirdly, since the training data is a small data set, data augmentation is very effective for the WER improvement, as indicated by System $\textrm{S}_2$, where speed perturbation (3x) is employed. 
Finally, we notice that we employ the ReLU activation function for mask estimate in the SE component of $\textrm{S}_4$,$\textrm{S}_5$,$\textrm{S}_6$ in Table \ref{tab:res}. However,in the $\textrm{S}_3$,$\textrm{S}_7$,we employ the Mish activation function for mask estimate

For the dual-channel MTJL (DC-MTJL) method, we are interested to see how the final WER result is affected by the clean data weighing factor $\gamma$ as indicated in Equation \ref{eqn:dual-loss}.
Table \ref{tab:dc-mtjl-res-ana} reports the WER results with 
different clean data weighting factor configuration, $\gamma$ in Equation (\ref{eqn:dual-loss}).

\begin{table}[ht]
    \centering
        \caption{WER~(\%) results for the dual-channel multitask-based joint learning method with different clean data weighting factor $\gamma$. }
    \label{tab:dc-mtjl-res-ana}
    \begin{tabular}{ccc}
    \toprule
    % \hline\hline
    System & Clean data weighing factor ($\gamma$) & WER (\%) \\
    % \hline\hline
    \midrule
    $\textrm{S}_1$     & 0.3  & 49.6 \\
    %\hline
    $\textrm{S}_2$ & 0.4 & 49.3\\
    %\hline
     $\textrm{S}_3$ & 0.5 & 49.9\\
     %\hline
      $\textrm{S}_4$ & 0.6 & 50.3\\
      %\hline
      $\textrm{S}_5$ & 0.7& \textbf{48.2}\\
        % \hline
         \bottomrule
    \end{tabular}
\end{table}
From Table \ref{tab:dc-mtjl-res-ana}, we observe that reasonable WER results can be achieved with $\gamma$ around 0.5. In our work, $\gamma=0.7$ yields the best WER. This suggests that one can obtain better recognition performance when clean data is combined to train the ASR system under very noise conditions.

Finally, as above mentioned, we attempted different activation function to estimate the mask in the front-end SE component. Based on the single channel multitask-based joint learn system as illustrated in Figure \ref{fig:multitask-joint-arch-01}, Table \ref{tab:activation-res} reports the performance comparison in detail.

\begin{table}[h]
    \centering
    \caption{WER (\%) results of using different activation function for mask estimation with single channel multitask-based joint learning configuration.}\label{tab:activation-res}
    \begin{tabular}{cc}
    \toprule
    % \hline\hline
     Activation type & WER (\%) \\
     % \hline\hline
      \midrule
      ReLU & \textbf{51.8}  \\
      %\hline
      Mish & 53.3 \\
      %\hline
      meta-ACON & 52.9 \\
      % \hline
      \bottomrule
    \end{tabular}
\end{table}
Table \ref{tab:activation-res} reveals that the ReLU activation function yields the best WER in our experiments.
%From 

\section{Conclusion}
\label{sec:con}
In this paper, we proposed a multitask-based joint learning framework for robust ASR over RATS radio communication speech data. Our discoveries lie in the following aspects. First, joint training can yield improved results, but keeping phase information is vital. Secondly, when joint training is combined with multitask recipe, further performance improvement can be achieved. Finally, since the target data is extremely noisy, training with the help of clean data is essential, which obtains the best WER reduction for the proposed method.

% 如果想跨栏显示图片，只需要添加在figure 这里添加*

%\section{REFERENCES}
%\label{sec:ref}
% \cite{C2}.\cite{Lamp86} 

% References should be produced using the bibtex program from suitable
% BiBTeX files (here: strings, refs, manuals). The IEEEbib.bst bibliography
% style file from IEEE produces unsorted bibliography list.
% -------------------------------------------------------------------------
\vfill\pagebreak
\clearpage
\bibliographystyle{IEEEbib}
\bibliography{refs}

\end{document}